\documentclass[floatfix,%
 aip,
 amsmath,amssymb,
 reprint,%
]{revtex4-1}

\usepackage{xcolor, color, soul}
\usepackage{graphicx}
\usepackage{dcolumn}
\usepackage{bm}

\usepackage[utf8]{inputenc}
\usepackage[T1]{fontenc}
\usepackage{mathptmx}
\usepackage{xcolor}
\usepackage{etoolbox}
\makeatletter
\def\@email#1#2{%
 \endgroup
 \patchcmd{\titleblock@produce}
  {\frontmatter@RRAPformat}
  {\frontmatter@RRAPformat{\produce@RRAP{*#1\href{mailto:#2}{#2}}}\frontmatter@RRAPformat}
  {}{}
}%
\makeatother
\usepackage{color,soul}
\begin{document}

\preprint{AIP/123-QED}

\title[]{A Theoretical Investigation of Magnetic Susceptibility Measurement of Diamagnetic Liquids Using a Mach-Zehnder Interferometer}
\author{David Shulman}
\affiliation{Department of Chemical Engineering, Ariel University, Ariel, Israel 407000}
\affiliation{Physics Department, Ariel University, Ariel 40700, Israel}
\email{davidshu@ariel.ac.il}

\date{\today}

\begin{abstract}
In this study, we present a novel method for measuring the magnetic susceptibility of liquids using a Mach-Zehnder interferometer. The proposed technique employs a ring magnet to deform the liquid, while a laser beam passes through the liquid to measure the resulting interference pattern. The deformation of the liquid, caused by the known magnetic field of the ring magnet, is used to calculate the magnetic susceptibility. We provide a comprehensive theoretical framework, including the relevant equations and models, for analyzing the data obtained using this method. We compare the Mach-Zehnder interferometer method with other established techniques, highlighting its advantages and disadvantages. Our findings indicate that the Mach-Zehnder interferometer technique offers high accuracy, sensitivity, and potential applications in characterizing magnetic properties of various liquid systems.
\end{abstract}


\maketitle 

\section{Introduction}
\subsection{Background and motivation}

Magnetic susceptibility \cite{landau1930diamagnetismus,landau1960electrodynamics} is a fundamental property of materials that characterizes their response to an applied magnetic field. The accurate measurement of magnetic susceptibility is essential for various applications in material science \cite{shulman2023measuring,keyser1989magnetic}, geophysics \cite{telford1990applied,hunt1995magnetic}, and medicine \cite{pranata2023optimizer}. Traditional techniques for measuring magnetic susceptibility include the Faraday balance method, the Gouy method \cite{saunderson1968permanent,morris1968faraday, splittgerber1971torsion}, and the vibrating-sample magnetometer \cite{lopez2018simple,guertin1974application}. However, these methods often require complex setups, large sample volumes, or intricate calibration procedures.

Interferometric techniques have been widely used for measuring small changes in physical properties, such as displacements \cite{michelson1887relative}, refractive indices \cite{korff1930sensitive}, and magnetic fields \cite{born2013principles,barker1972laser}. Among the various interferometers, the Mach-Zehnder \cite{ludwig1891neuer,mach1892ueber} and Michelson \cite{michelson1887relative} interferometers have gained significant popularity due to their sensitivity and versatility. Recently, a novel technique exploiting the Moses effect to measure the magnetic susceptibility of diamagnetic liquids was proposed \cite{SHULMAN2023}. This technique relies on the deformation of the liquid-vapor interface by a steady magnetic field and measures the deformation using an optical method.

In this study, we propose a novel method to measure the magnetic susceptibility of liquids using a Mach-Zehnder interferometer, which has the potential for high precision, simplicity, and low sample volume requirements.
\subsection{Magnetic susceptibility}

Magnetic susceptibility ($\chi$) is a dimensionless quantity that describes the extent to which a material is magnetized in response to an applied magnetic field. Materials can be classified into three categories based on their magnetic susceptibility: diamagnetic, paramagnetic, and ferromagnetic \cite{landau1930diamagnetismus,landau1960electrodynamics}. Diamagnetic materials have negative susceptibilities, while paramagnetic and ferromagnetic materials exhibit positive susceptibilities. Magnetic susceptibility is a crucial property for understanding the magnetic behavior of materials, and it is essential for various applications such as magnetic resonance imaging \cite{beuf1996magnetic}, magnetic separation \cite{oberteuffer1974magnetic}, and the study of magnetic materials.
\subsection{Mach-Zehnder interferometer}

The Mach-Zehnder interferometer \cite{ludwig1891neuer,mach1892ueber} is an optical device that splits a coherent light beam into two separate paths and recombines them to form an interference pattern. It consists of two beam splitters and two mirrors arranged in such a way that the recombined beams interfere constructively or destructively, depending on the phase difference between the two paths. This phase difference can be caused by changes in the optical path length or the refractive index of the materials through which the beams pass. The Mach-Zehnder interferometer has been widely used for measuring small changes in refractive index, temperature, pressure, and other physical quantities.

\section{Theory}

\subsection{Interference in Mach-Zehnder Interferometer}
The Mach-Zehnder interferometer is an optical device that uses two beam splitters to split an incoming laser beam into two separate beams, which are then recombined to form an interference pattern. The principle of interference in a Mach-Zehnder interferometer can be described as follows:

\begin{enumerate}
\item An incoming collimated light beam is divided into two beams, the sample beam (SB) and the reference beam (RB), by a beam splitter (BS1).
\item The SB is reflected by mirrors M1 and M2, while the RB takes a separate path.
\item Both beams are recombined at a second beam splitter (BS2), creating an interference pattern as a result of the phase difference between the two beams due to their separate paths through the interferometer.
\end{enumerate}

\subsection{Magnetic Field-induced Deformation of Liquid}
When a magnetic field is applied to a diamagnetic liquid, the liquid's magnetic susceptibility causes it to deform in the presence of the field. The deformation depends on the liquid's magnetic susceptibility and the strength of the magnetic field. In our setup, a ring magnet is used to create a magnetic field, and the deformation of the liquid is observed as a change in the interference pattern.

\subsection{Relation between Interference Fringes and Change in Thickness $\Delta z$}

First, we relate the phase shift $\Delta \phi$ in the interferometer to the change in the optical path length $\Delta L$.

\begin{equation}
\Delta \phi = \frac{2 \pi}{\lambda} \Delta L
\end{equation}

where $\lambda$ is the wavelength of the laser and $\Delta L$ is the change in the optical path length.

Next, we relate the change in the optical path length $\Delta L$ to the change in thickness $\Delta z$. The change in the optical path length is equal to the product of the change in thickness and the difference in refractive indices $(n_{\text{liquid}} - n_{\text{air}})$.

\begin{equation}
\Delta L = (n_{\text{liquid}} - n_{\text{air}})  \Delta z
\end{equation}

Now, we relate the number of fringes $\Delta n$ to the change in phase shift $\Delta \phi$. The change in phase shift is equal to the product of the number of fringes and $2\pi$.

\begin{equation}
\Delta \phi = 2 \pi \Delta n
\end{equation}

Combining the three equations above, we obtain the following relationship between the number of fringes $\Delta n$ and the change in thickness $\Delta z$:

\begin{equation}
\Delta n = \frac{(n_{\text{liquid}} - n_{\text{air}}) }{\lambda} \Delta z
\end{equation}

This equation relates the number of fringes $\Delta n$ to the change in thickness $\Delta z$ of the liquid, with $(n_{\text{liquid}} - n_{\text{air}})$ being the difference in refractive indices and $\lambda$ the wavelength of the laser. By measuring the number of fringes and knowing the other parameters, we can calculate the change in thickness $\Delta z$ of the liquid.
\subsection{Profile of Deformed Liquid Surface and Magnetic Susceptibility Calculation}

The profile of the deformed liquid surface can be derived from the Laplace equation. A detailed derivation of this relationship can be found in the Appendix.

To calculate the magnetic susceptibility, we first measure the number of fringes and use the previously derived relationship:

\begin{equation}
\Delta n = \frac{(n_{\text{liquid}} - n_{\text{air}})}{\lambda} \Delta d
\end{equation}

By rearranging the equation and knowing the refractive indices and the wavelength of the laser, we can calculate the change in thickness $\Delta d$:

\begin{equation}
\Delta d = \frac{\lambda \Delta n}{(n_{\text{liquid}} - n_{\text{air}})}
\end{equation}

\begin{widetext}
\begin{figure*}[!t]
\normalsize
  \begin{minipage}{0.48\textwidth}
     \centering
     \includegraphics[width=1\linewidth]{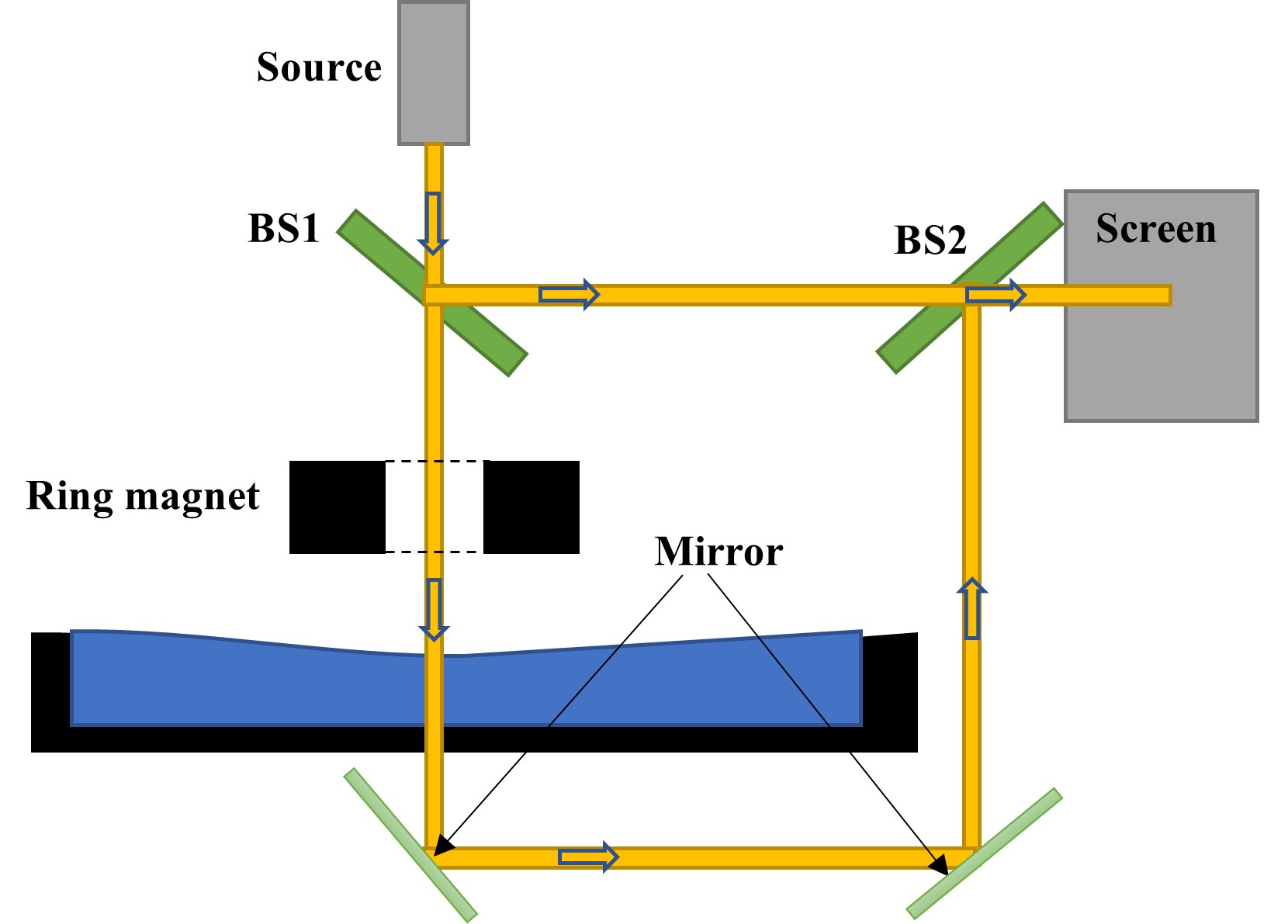}
     \caption{The schematic outline of the experiment setup}\label{Setup_N}
  \end{minipage}\hfill
  \begin{minipage}{0.48\textwidth}
     \centering
     \includegraphics[width=1\linewidth]{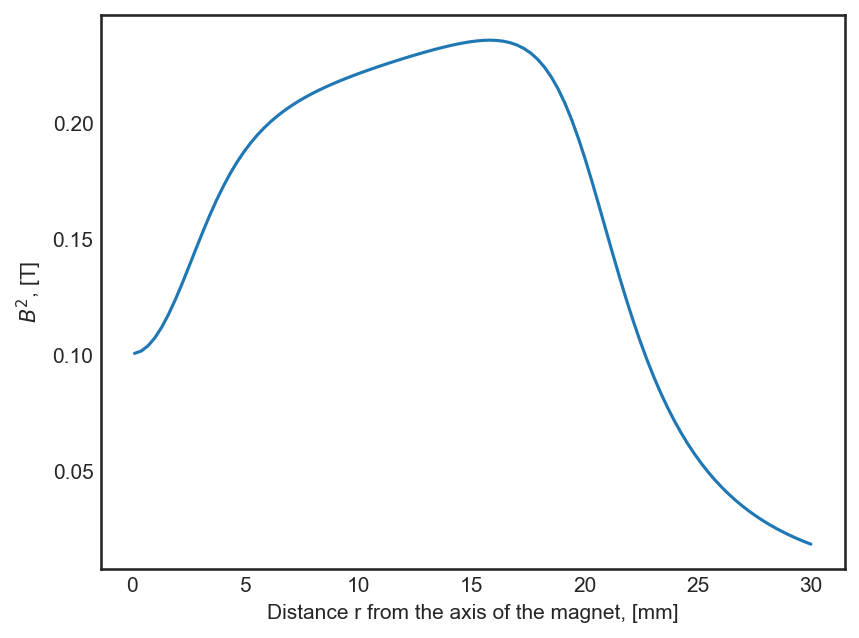}
     \caption{Square of the magnetic field $B^2$ as a function of radial distance from the center of a ring magnet with length 30 mm, outer diameter 40 mm, and inner diameter 5 mm, for a distance $h$ between the magnet and the liquid surface of 3 mm.}\label{B_squar}
  \end{minipage}
\hrulefill
\vspace*{4pt}  
  \begin{minipage}{0.48\textwidth}
     \centering
\includegraphics[width=\linewidth]{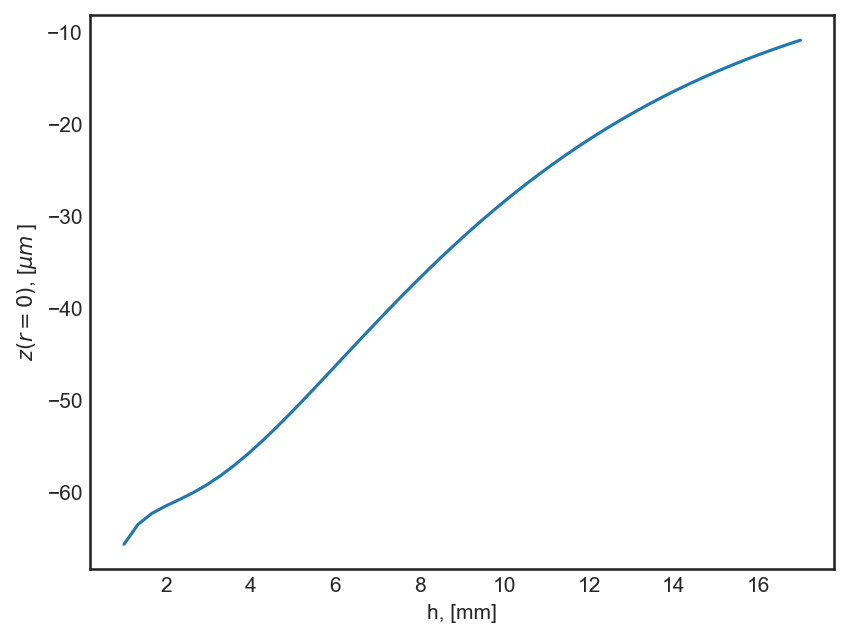}
\caption{This graph depicts the correlation between the maximum depth of the deformed liquid and the distance between the magnet and liquid surface.}
\label{fig:z_maxB}
  \end{minipage}\hfill
  \begin{minipage}{0.48\textwidth}
     \centering
\includegraphics[width=\linewidth]{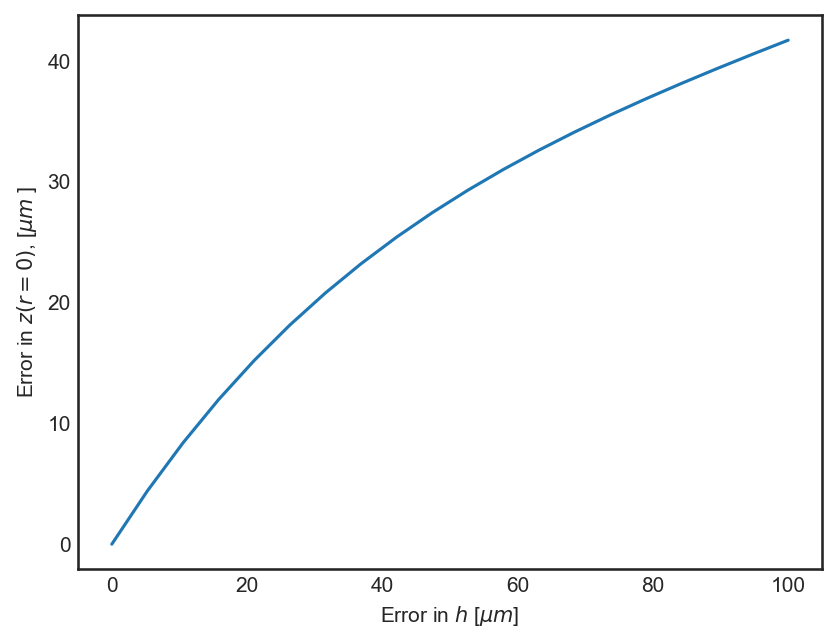}
\caption{The graph displays the relationship between the error in the distance between the magnet and liquid surface and the resulting error in the maximum depth of the deformed liquid profile.}
\label{fig:z_maxBEr}
  \end{minipage}
\vspace*{4pt} 
  \begin{minipage}{0.48\textwidth}
     \centering
\includegraphics[width=\linewidth]{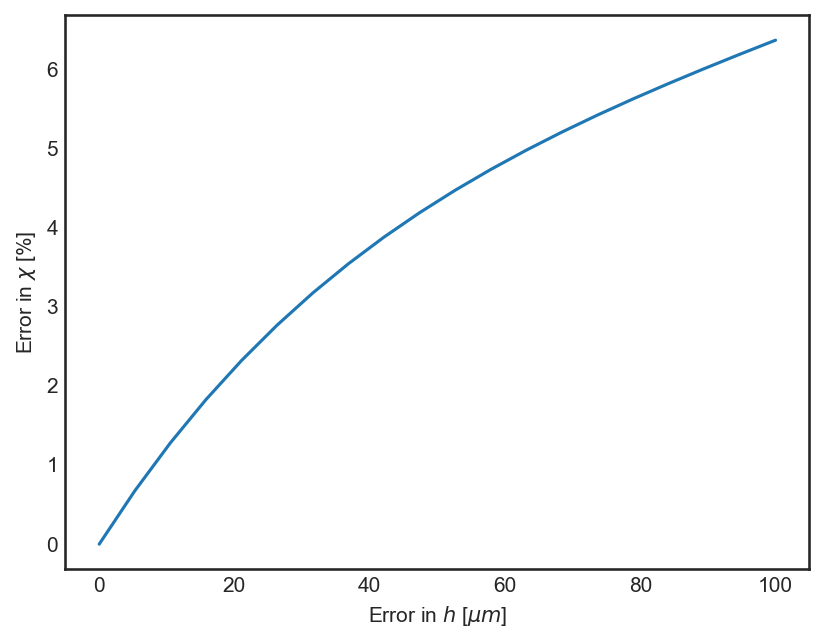}
\caption{Graph showing the error in magnetic susceptibility $\chi$ measurements as a function of the error in the distance $h$ between the magnet and the liquid surface. }
\label{fig:chiEr}
  \end{minipage}\hfill
\vspace*{4pt} 
\end{figure*}
\end{widetext}

The magnetic field model used for the calculation is presented in the Appendix. The model for the $B^2$ of the magnetic field can be chosen as the solution of the Biot-Savart law for the ideal solenoid, as described in the appendix \cite{Shulman2023Mag}. The relationship between the square of the magnetic field $B^2$ and the radial distance $r$ from the center of a ring magnet is illustrated in Fig. (\ref{B_squar}). Using the derived expression for the magnetic susceptibility Eq.(\ref{chi}) and the magnetic field model, we can calculate the magnetic susceptibility of the liquid.

Before conducting measurements on unknown liquids, it is essential to calibrate the apparatus using a known liquid with a known magnetic susceptibility. By performing calibration measurements, we can ensure the accuracy and reliability of the calculated magnetic susceptibility for the unknown liquid samples.
\section{Measurement Procedure}

\subsection{Experimental Setup}

The experimental setup for measuring the magnetic susceptibility of liquids using a Mach-Zehnder interferometer consists of the following components, see Fig. (\ref{Setup_N}):

\begin{itemize}
\item A He-Ne laser with a wavelength of 633 nm to generate a collimated beam of light.
\item A beam splitter (BS1) to split the laser beam into two paths: the sample beam (SB) and the reference beam (RB).
\item Mirrors M1 and M2 to direct the SB towards the second beam splitter (BS2).
\item A container filled with the diamagnetic liquid and a ring magnet placed at a fixed distance $h$ above the container.
\item A micrometer stage to move the ring magnet up and down, causing changes in the thickness of the diamagnetic liquid.
\item A second beam splitter (BS2) to recombine the SB and RB, creating an interference pattern.
\end{itemize}
\subsection*{Ring Magnet and Micrometer Stage}

A ring magnet is used to generate a static magnetic field, which deforms the liquid sample in the container. The magnetic field strength and uniformity are critical for accurate measurements of the magnetic susceptibility of the diamagnetic liquid.

The ring magnet is mounted on a micrometer stage, which allows for precise control of the magnet's position relative to the liquid container. By moving the ring magnet up and down using the micrometer stage, the thickness of the liquid layer can be altered, resulting in a change in the optical path length and thus the interference pattern observed on the detector or screen.

\subsection*{Liquid Container}

The liquid container holds the diamagnetic liquid sample under investigation. It should be made of a transparent material with a high optical quality, such as quartz or borosilicate glass, to minimize distortion of the light beam passing through the container. The container should also have a flat surface to ensure a stable interference pattern.

\subsection{Procedure Steps}

The following steps outline the procedure for measuring the magnetic susceptibility of liquids using a Mach-Zehnder interferometer:

\begin{enumerate}
\item Place a permanent ring magnet at a fixed distance $h$ above a container filled with diamagnetic liquid.
\item Use a He-Ne laser with a wavelength of 633 nm to generate a collimated beam of light.
\item Split the beam into two paths (SB and RB) by a beam splitter (BS1). Direct the SB towards the diamagnetic liquid.
\item Pass the SB through the ring magnet and subsequently through the liquid and the container. Reflect the beam using mirrors M1 and M2.
\item Recombine the SB and RB at the second beam splitter (BS2) and observe the interference fringes.
\item Use a micrometer stage to move the ring magnet up and down, causing changes in the thickness of the diamagnetic liquid. Fig. (\ref{fig:z_maxB}) displays the relationship between the maximum depth of the deformed liquid profile and the distance between the magnet and liquid surface.
\item Count the number of fringes observed as the ring magnet is moved, and calculate the change in thickness and the magnetic susceptibility of the liquid from the number of fringes.
\end{enumerate}

\subsection{Calibration of the Apparatus}

Calibration of the apparatus is an essential step to ensure accurate measurements of the magnetic susceptibility of liquids. The calibration process is as follows:

\begin{enumerate}
\item Fill the container with a known liquid with a known magnetic susceptibility.
\item Measure the number of fringes observed as the ring magnet is moved using the micrometer stage.
\item Using the known magnetic susceptibility and the measured number of fringes, verify the accuracy of the derived equation relating the number of fringes to the change in thickness of the liquid and the magnetic field model of the permanent magnet.
\item Adjust the apparatus or the experimental parameters, if necessary, to improve the accuracy of the measurements.
\end{enumerate}

After calibrating the apparatus, you can proceed with the measurements of the magnetic susceptibility for unknown liquid samples.
\section{Results and Discussion}

\subsection{Simulation Results}

Using the derived equations and the experimental setup, we performed numerical simulations to analyze the magnetic susceptibility measurements of various liquids. The results of these simulations provide insights into the accuracy, sensitivity, and precision of the Mach-Zehnder interferometer technique.

To assess the impact of height errors on the accuracy of magnetic susceptibility measurements, we generated a plot that illustrates the relationship between the error in the distance between the magnet and liquid surface and the corresponding error in the maximum depth of the deformed liquid profile, as shown in Fig. (\ref{fig:z_maxBEr}). 

By utilizing the exact equation for the deformed profile, we are able to simulate and estimate the error in the magnetic susceptibility measurement as a function of the error in the distance between the magnet and the liquid surface, as shown in Fig. (\ref{fig:chiEr}). From the graph, we can observe that the usual accuracy in distance measurement for $h$, which is about $\pm 10\ \mu m$, yields an accuracy in $\chi$ of about $1\%$, which is an acceptable level of error for many practical applications.
\subsection{Comparison with Other Measurement Techniques}

Here, we compare the Mach-Zehnder interferometer method with the method for measuring magnetic susceptibility of diamagnetic liquids exploiting the Moses effect \cite{SHULMAN2023}. The pros and cons of both methods are discussed below:

\begin{enumerate}
\item Using the Moses Effect \cite{SHULMAN2023} to measure the maximal slope of the liquid/air interface deformation:
Pros:
\begin{itemize}
    \item Simpler experimental setup: The Moses Effect method \cite{SHULMAN2023} requires fewer optical components and is generally easier to set up.
\end{itemize}

Cons:
\begin{itemize}
    \item Less accurate: The method relies on measuring the maximal slope of the liquid/air interface, which can be affected by various optical and surface-related side effects, resulting in less accurate measurements.
\end{itemize}

\item Using a Mach-Zehnder Interferometer to measure the change in thickness of the deformed liquid:

Pros:
\begin{itemize}
    \item More direct measurement of magnetic susceptibility: The Mach-Zehnder Interferometer method measures the change in thickness of the deformed liquid directly, avoiding the side optical effects that can affect the accuracy of the Moses Effect method \cite{SHULMAN2023}.
    \item Simpler mathematical calculations: The calculations involved in analyzing the data from the Mach-Zehnder Interferometer method are generally more straightforward than those required for the Moses Effect method \cite{SHULMAN2023}.
    \item Better accuracy: The Mach-Zehnder Interferometer method typically provides higher accuracy in measuring magnetic susceptibility compared to the Moses Effect method.
\end{itemize}

Cons:
\begin{itemize}
    \item More complex experimental setup: The Mach-Zehnder Interferometer requires more optical components and precise alignment, making the experimental setup more complicated than the Moses Effect method.
\end{itemize}
\end{enumerate}

\subsection{Limitations and Sources of Error}

Despite its potential advantages, the Mach-Zehnder interferometer method has certain limitations and sources of error that can affect the accuracy and reliability of the measurements:

\begin{itemize}
\item Systematic errors due to the misalignment of optical components, such as beam splitters and mirrors.
\item Thermal fluctuations causing variations in the refractive index of the liquid, leading to errors in the measured magnetic susceptibility.
\item Imperfections in the magnetic field generated by the ring magnet, which can affect the deformation of the liquid surface and, consequently, the measured magnetic susceptibility.
\item Errors in the calibration process, which can propagate through the measurements and calculations.
\end{itemize}

\subsection{Potential Applications and Future Work}

The Mach-Zehnder interferometer method for measuring the magnetic susceptibility of liquids has several potential applications, including the characterization of novel materials, the study of magnetic properties in biological systems, and the investigation of magnetic phenomena in soft condensed matter.

Future work could focus on addressing the limitations and sources of error mentioned in the previous subsection, as well as exploring the potential of the Mach-Zehnder interferometer method for measuring magnetic susceptibility in other systems, such as suspensions, emulsions, and composite materials. Additionally, further research could explore the integration of this method with other optical techniques, to provide a comprehensive characterization of the magnetic properties and molecular structure of the samples under investigation.

\section{Conclusion}

In this study, we presented a theoretical investigation of the magnetic susceptibility measurement of liquids using a Mach-Zehnder interferometer. We derived the equations and models required for the analysis and provided a detailed description of the experimental setup and procedure. The method involves deforming a liquid sample with a ring magnet and analyzing the resulting interference pattern using a Mach-Zehnder interferometer to calculate the magnetic susceptibility of the liquid.

The simulation results demonstrated the potential of this technique for measuring the magnetic susceptibility of various liquids with high accuracy and sensitivity. We also compared the Mach-Zehnder interferometer method with other established techniques, highlighting its advantages and disadvantages. Despite some limitations and sources of error, the method offers promising applications in the characterization of novel materials, biological systems, and soft condensed matter.

Future work should focus on addressing the limitations and errors identified in the study, exploring the potential of the Mach-Zehnder interferometer method for other systems, and integrating it with other optical techniques to provide a comprehensive characterization of the magnetic properties and molecular structure of samples. Overall, the Mach-Zehnder interferometer method has the potential to become a valuable tool in the investigation of magnetic susceptibility in various liquid systems.




\newpage
\appendix
\begin{widetext}
\section{Using the Exact Solution of the Laplace Equation for Deformed Liquid Profile to Improve the Measurement of Magnetic Susceptibility}

The Young-Laplace equation, which describes the equilibrium shape of a liquid surface in response to external forces, including magnetic forces. The left-hand side of the equation represents the gravitational and surface tension forces, while the right-hand side represents the magnetic force, see Ref. 

\begin{equation}
\frac{\partial ^{2}z}{\partial r^{2}}+\frac{1}{r}\frac{\partial z}{\partial r%
}-\frac{\rho g}{\gamma }z=\frac{\chi B^{2}\left( r,h\right) }{2\mu
_{0}\gamma }  \label{z_equation}
\end{equation}
The solution to this equation gives the profile of the deformed liquid surface: 
\begin{equation}
z\left( r,h\right) =-\left[ \int_{0}^{r}\frac{\chi B^{2}\left( r^\prime,h\right) }{%
2\mu _{0}\gamma }I_{0}\left( \lambda _{c}^{-1}r^\prime\right) r^\prime dr^\prime\right]
K_{0}\left( \lambda _{c}^{-1}r\right)-\left[ \int_{r}^{\infty }\frac{\chi
B^{2}\left( r^\prime,h\right) }{2\mu _{0}\gamma }K_{0}\left( \lambda
_{c}^{-1}r^\prime\right) r^\prime dr^\prime\right] I_{0}\left( \lambda _{c}^{-1}r\right) \label{z}
\end{equation}

where $\gamma$ is the surface tension of the liquid, $h$ is the separation between the magnet and liquid/vapor interface, $I_0$ and $K_0$ are modified Bessel functions of the first and second kind, respectively. The interplay between the gravity and the surface tension is quantified by the capillary length, denoted $\lambda_c$. 

The magnetic susceptibility can be determined using Eq. (\ref{z}) as follows:

\begin{equation}
\chi =-z\left( r,h\right)\left(  \left[ \int_{0}^{r}\frac{ B^{2}\left( r^\prime,h\right) }{%
2\mu _{0}\gamma }I_{0}\left( \lambda _{c}^{-1}r^\prime\right) r^\prime dr^\prime\right]
K_{0}\left( \lambda _{c}^{-1}r\right)+\left[ \int_{r}^{\infty }\frac{
B^{2}\left( r^\prime,h\right) }{2\mu _{0}\gamma }K_{0}\left( \lambda
_{c}^{-1}r^\prime\right) r^\prime dr^\prime\right] I_{0}\left( \lambda _{c}^{-1}r\right)\right)^{-1} \label{chi}
\end{equation}

The model of the magnetic field used in this investigation is the solution of the Biot-Savart law for the ideal solenoid \cite{Shulman2023Mag}:

\begin{equation}
B_{r}\left( r,h\right) =B_{0}\int_{0}^{\pi /2}d\psi \left( \cos ^{2}\psi
-\sin ^{2}\psi \right) \left\{ \frac{\alpha _{+}}{\sqrt{\cos ^{2}\psi
+k_{+}^{2}\sin ^{2}\psi }}-\frac{\alpha _{-}}{\sqrt{\cos ^{2}\psi
+k_{-}^{2}\sin ^{2}\psi }}\right\}
\end{equation}

\begin{equation}
B_{z}\left( r,h\right) =\frac{B_{0}a}{r+a}\int_{0}^{\pi /2}d\psi \left( 
\frac{\cos ^{2}\psi +\tau \sin ^{2}\psi }{\cos ^{2}\psi +\tau ^{2}\sin
^{2}\psi }\right) \left\{ \frac{\beta _{+}}{\sqrt{\cos ^{2}\psi
+k_{+}^{2}\sin ^{2}\psi }}-\frac{\beta _{-}}{\sqrt{\cos ^{2}\psi
+k_{-}^{2}\sin ^{2}\psi }}\right\}
\end{equation}

\begin{equation*}
\alpha _{\pm }=\frac{a}{\sqrt{h_{\pm }^{2}+\left( r+a\right) ^{2}}},\ \ \ \
\ \ \ \ \ \beta _{\pm }=\frac{h_{\pm }}{\sqrt{h_{\pm }^{2}+\left( r+a\right)
^{2}}}\ \ \ \ \ \ 
\end{equation*}

\begin{equation*}
h_{+}=h,\ \ \ h_{-}=h-2b,\ \ \ \ \tau =\frac{a-r}{a+r}
\end{equation*}

\begin{equation*}
k_{\pm }=\sqrt{\frac{h_{\pm }^{2}+\left( a-r\right) ^{2}}{h_{\pm
}^{2}+\left( a-r\right) ^{2}}}
\end{equation*}

where $a$ is the radius and $2b$ is the length of the solenoid; $\left(r,\ \varphi,\ h\right)$ are the cylindrical coordinates with the origin at the center of the solenoid; $n$ – is the number of turns per unit length. To obtain the equations in the current form, we have also introduced the following integration variable change: $2\psi\equiv\pi-\varphi$. 

$B_0$ is a coefficient of the magnetic field generated by an ideal solenoid with $n$ turns per unit length and carrying a current $I$. The formula for $B_0$ is given by:
\begin{equation*}
B_0=\frac{\mu_0}{\pi}nI,
\end{equation*}
where $\mu_0$ is the magnetic constant (also known as the permeability of free space), which is equal to $4\pi\times 10^{-7}$ tesla meter per ampere (T m/A). The value of $B_0$ depends on the geometry of the solenoid and the current flowing through it. It is an important parameter for calculating the magnetic field and the magnetic susceptibility of a diamagnetic liquid using the proposed method in this manuscript.
When using a permanent magnet instead of an ideal solenoid, the coefficient $B_0$ cannot be directly calculated and must be estimated experimentally. Therefore, the experimental determination of $B_0$ is necessary to accurately calculate the magnetic susceptibility using the proposed method.

Before taking measurements, it is necessary to calibrate the apparatus using a known liquid.

When using the exact solution for the deformed liquid profile (Eq. \ref{z}) to determine the magnetic susceptibility, the following steps can be taken:
\begin{enumerate}
  \item Measure the number of fringes observed for a given change in thickness between two positions of the magnet, $h_0$ and $h_1$.
  \item Calculate the change in thickness of the liquid film, $\Delta z$.
  \item Extract $\chi$ from $\Delta z$, using Eq. \ref{chi}.
  \item Repeat the experiment and take the mean of the calculated values to obtain a more precise estimation of the magnetic susceptibility.
\end{enumerate}
\end{widetext}

\section*{Conflict of Interest}

The author has no conflicts to disclose.
\begin{acknowledgments}
I would like to express my deep gratitude to Professor Meir Lewkowicz and Professor Edward Bormashenko, my research supervisors, for their patient guidance, enthusiastic encouragement, and useful critiques
\end{acknowledgments}

\bibliography{aipsamp}
\end{document}